\documentclass{ifacconf}

\usepackage{graphicx}      % include this line if your document contains figures
\usepackage{natbib}        % required for bibliography

\usepackage{amsmath}
\usepackage{amssymb}
\usepackage{amsfonts}
\usepackage{xcolor}
\usepackage{physics}
\usepackage[per-mode=symbol]{siunitx}
\usepackage{optidef}
\usepackage{algorithm}
\usepackage{algpseudocode}
\usepackage{microtype}
\usepackage{scalerel}
\usepackage{tikz}

\DeclareMathOperator{\col}{col}
\DeclareMathOperator{\specialo}{SO}
\DeclareMathOperator{\Int}{Int}
 
\graphicspath{{img/}}
\newcommand{\executeiffilenewer}[3]{%
	\ifnum\pdfstrcmp{\pdffilemoddate{#1}}%
	{\pdffilemoddate{#2}}>0%
	{\immediate\write18{#3}}\fi%
}

\begin{document}
\begin{frontmatter}

\title{Task-Priority Control of Redundant Robotic Systems using Control Lyapunov and Control Barrier Function based Quadratic Programs\thanksref{footnoteinfo}} 
% Title, preferably not more than 10 words.

\thanks[footnoteinfo]{This research was partly funded by the Research Council of
Norway through the Centres of Excellence funding scheme, project
No. 223254 NTNU AMOS.}

\author[First]{Erlend A. Basso}  
\author[First]{Kristin Y. Pettersen} 

\address[First]{Centre for Autonomous Marine Operations and Systems (NTNU AMOS), 
   Department of Engineering Cybernetics, Norwegian University of Science and Technology, Trondheim, Norway (e-mail: erlend.a.basso@ntnu.no).}

\begin{abstract}                % Abstract of not more than 250 words.
  This paper presents a novel task-priority control framework for redundant robotic systems based on a hierarchy of control Lyapunov function (CLF) and control barrier function (CBF) based quadratic programs (QPs). The proposed method guarantees strict priority among different groups of tasks such as safety-related, operational and optimization tasks. Moreover, a soft priority measure in the form of penalty parameters can be employed to prioritize tasks at the same priority level. As opposed to kinematic control schemes, the proposed framework is a holistic approach to control of redundant robotic systems, which solves the redundancy resolution, dynamic control and control allocation problems simultaneously. Numerical simulations of a hyper-redundant articulated intervention autonomous underwater vehicle (AIAUV) is presented to validate the proposed framework.
\end{abstract}


\end{frontmatter}
%===============================================================================
\section{Introduction}
A robotic system is kinematically redundant when it has more degrees of freedom (DOFs) than those strictly required to execute a given task. Tasks are often divided into groups according to their priority, such as safety-related tasks, operational tasks and optimization tasks. Kinematic redundancy enables additional tasks to be executed simultaneously by utilizing the redundant DOFs of the system. However, care must be taken when resolving kinematic redundancy. In particular, if compatibility between two or more tasks cannot be guaranteed at all time, then lower-priority tasks may prevent convergence of higher-priority tasks. Since tasks at the lowest-priority level are often added to ensure joint-space stability of a robotic system, they are inherently incompatible with mission-related tasks such as end-effector control. Consequently, redundancy must be resolved by ensuring strict-priority between tasks, i.e. that lower-priority tasks have no effect on the execution of higher-priority tasks.

% strict priority between tasks should be ensured such that lower-priority tasks 
%  Redundancy should be resolved such that lower-priority tasks do not affect the execution of higher-priority tasks.
% It is useful to divide control tasks into three groups, safety-related tasks, operational tasks and optimization tasks, arranged by decreasing priority \citep{dilillo2018}. 

Kinematic task-priority control is a redundancy resolution method introduced in \cite{Hanafusa1981}, developed in \cite{Nakamura1987} and generalized to any number of priority levels in \cite{Siciliano1991}. This control approach decouples the controller into a kinematic and dynamic controller, and has been successfully implemented on a number of robotic systems. The framework was extended to support tasks described by sets or inequalities in \cite{Moe2016}, \cite{Simetti2016} and \cite{Kanoun2011}. These kinematic control approaches all resolve redundancy at the velocity level by generating velocity references for some dynamic controller to follow. An immediate drawback is that acceleration references cannot be included, resulting in poorer tracking accuracy. 

% However, the original framework is prone to algorithmic singularities and the singularity robust task-priority framework \citep{Chiaverini1997}, \citep{Antonelli2010} executes lower-priority tasks poorly.

% Alternatively, by resorting to operational space control \citep{Khatib1987}, redundancy can be resolved at the torque level in a prioritized manner for an arbitrary number of tasks \citep{Sentis2004}. 

% Alternatively, operational space control \citep{Khatib1987} transforms the dynamic equations of motion of a robotic system from joint space into the operational space, also known as task space. Although originally introduced for two tasks, the scheme was extended to a task-priority framework with an arbitrary number of tasks in \citep{Sentis2004}. task-priority operational space control resolves redundancy at the torque level by employing null-space operators in the control law. These null-space operators ensure that torques generated by lower-priority tasks do not generate accelerations or forces affecting the task dynamics of higher-priority tasks. 
% \todo[color=green!40,inline]{operational space task-priority, shortcomings wrt set-based tasks}
Operational space control \citep{Khatib1987} is a holistic approach that assigns joint torques directly by transforming the equations of motion from joint space into the operational space (also known as task space). Although it was mainly introduced for non-redundant systems, a dynamically consistent null space operator was defined in \cite{Khatib1987}, that allowed two operational space tasks to be defined and controlled simultaneously. In \cite{Sentis2004}, the scheme was extended to a task-priority framework with an arbitrary number of tasks by generalizing the dynamically consistent null space operator from \cite{Khatib1987} to an arbitrary number of priority levels. These null space operators ensure that torques generated by lower-priority tasks do not generate accelerations that affect the task dynamics of higher-priority tasks. The operational space framework was extended to include set-based tasks, i.e. control objectives with a range of valid values in \cite{4926146}, but this approach does not scale well for systems with a high number of DOFs. 
% \todo[inline]{earlier work on CLF-CBF... $\rightarrow$ CLF-CBF task-priority framework} 

Control Lyapunov functions (CLFs) extend Lyapunov theory to systems with inputs and have become an essential part of nonlinear control design after the pioneering work in \cite{Artstein1983,Sontag1983,Sontag1989}.  The CLF concept was extended to rapidly exponentially stabilizing control Lyapunov functions (RES-CLFs) in \cite{AmesRES}, which achieve exponential convergence at a controllable rate. Through CLFs or RES-CLFs, the control designer is free to chose among an infinite number of controllers. An important example is the point-wise minimum norm controller \citep{Freeman1996,Petersen1987}, which selects the control value of minimum norm from all control values rendering the time derivative of the CLF negative definite. The point-wise minimum norm controller has a closed-form solution since it is the solution to a quadratic program (QP) with only one inequality constraint. This QP can be augmented with control input saturation limits and other control input constraints, at the expense of a closed-form solution \citep{Galloway2015}. For redundant robotic systems, two control tasks can be satisfied simultaneously by defining CLFs for each task and finding a control input that minimizes some quadratic objective function while ensuring that the time derivatives of the CLFs are negative definite \citep{Ames2013}. However, strict priority between tasks cannot be ensured.

Barrier functions have been used extensively in constrained optimization \citep{Forsgren2002, StephenBoyd2019}, and have motivated the concept of barrier certificates for safety-critical control. Barrier certificates were introduced as a tool for proving forward invariance of sets \citep{Prajna2004,Prajna2006}. Since these sets often encoded safety-related objectives, proving invariance of a safe set implies that the system will remain safe, as long as you start safe. These barrier certificates tend to infinity as the state tends to the boundary of the safe set, and in order to obtain safety guarantees beyond the boundary of the safe set, various Lyapunov-like approaches have been proposed such as \citep{Tee2009}, where a positive definite barrier certificate is employed as a barrier Lyapunov function. However, these conditions are overly conservative since the positive definiteness property enforces the invariance of every level set, and not just the safe set of the set-based task in question.

Barrier certificates were extended to systems with inputs by introducing the first notion of a control barrier function (CBF) in \cite{Wieland2007}. These control barrier functions were combined with control Lyapunov functions in \cite{Romdlony2014}, and further improved in \cite{Romdlony2016} to establish conditions for so-called control Lyapunov-barrier functions, which jointly guarantee safety and stability. However, these conditions were shown to be too restrictive, and subsequently relaxed in \cite{Ames2014,Ames2017}, which extended control barrier functions to the entire safe set, and thus enabling controller synthesis through optimization-based methods \citep{Ames2019}. In particular, the CLF-based QPs in \cite{Ames2013} and \cite{Galloway2015} could be augmented with CBFs to ensure stability and safety \citep{Ames2014,Ames2017}. CBFs were generalized to exponential control barrier functions (ECBFs) in \cite{Nguyen2016}, which enforce forward invariance of set-based tasks with a higher relative degree. 

The main contribution of this paper is a novel dynamic task-priority framework for an arbitrary number of equality- and set-based control tasks encoded by CLFs and CBFs, where equality and set-based tasks are control objectives that should be driven to a desired value and kept within a desired set, respectively. The framework builds on the CLF-based QP proposed for two equality tasks in \cite{Ames2013} by extending it to an arbitrary number of equality tasks, unifying CLFs with CBFs via QPs to support set-based tasks as done in \cite{Ames2014,Ames2017}, and establishing any number of priority levels through a hierarchy of QPs. An important feature of this approach is that it yields strict priority between tasks at different priority levels, in the sense that tasks at lower-priority levels have no effect on the execution of tasks at higher-priority levels. The inclusion of set-based tasks at the dynamic level is a key novelty within task-priority control, which has to the best of our knowledge only been accounted for in \cite{4926146}.

The proposed scheme represents a holistic control approach since the QPs can be formulated in terms of the actuator inputs, instead of the commanded forces and torques. Consequently, the proposed framework also solves the control allocation problem. For task-priority control of robotic systems where computation of the actuator inputs from the commanded forces and torques is non-trivial, the unification of redundancy resolution and control allocation is a key advantage because strict priority between tasks can be ensured at all times.
% In contrast to redundancy resolution schemes such as kinematic or operational space control that decouple dynamic control and control allocation, the unification of redundancy resolution and control allocation ensures that strict priority between tasks can be guaranteed at all times. For redundancy resolution schemes that decouple dynamic control and control allocation, strict priority between tasks cannot be guaranteed. 
In contrast, redundancy resolution schemes that decouple dynamic control and control allocation, such as kinematic or operational space control, provide no a priori guarantee that the commanded forces and torques computed by the dynamic controller can be exactly allocated, since the commanded forces and torques are typically computed with no regard to physical actuator limits, rate constraints, or singularities of the actuator configuration matrix. If the commanded forces and torques cannot be exactly allocated, the forces and torques are usually allocated to actuator inputs by minimizing the allocation error \citep{JOHANSEN20131087}, which is performed independently of the redundancy resolution algorithm. As a result, strict priority is lost and tasks become coupled whenever exact allocation is infeasible.
 
% The framework is a holistic approach to robot control that is especially useful for redundant robots that require control allocation, since the QPs can be formulated in terms of the actual control input, and thereby solving the redundancy resolution, dynamic control and control allocation problems simultaneously.

% The method supports both equality- and set-based tasks represented by CLFs and CBFs, which are control objectives that should be driven to a desired value and kept within a desired set described by inequalities, respectively.
% In contrast to the CLF based QP proposed for two tasks in \cite{Ames2013}, the proposed framework establishes strict priority levels by utilizing a hierarchy of QPs, where tasks at a lower-priority level have no effect on the execution of tasks at higher-priority levels. Set-based tasks are included by employing CBFs or ECBFs 
% The framework builds on the work in \cite{Ames2013,Ames2014,Nguyen2016,Ames2019}. In particular, the CLF-based QP proposed for two tasks in \cite{Ames2013} is extended to an arbitrary number of tasks and combined with CBFs 

% This framework is especially useful for redundant robotic systems that require control allocation, since the control allocation and redundancy resolution problems can be solved simultaneously.

This paper is organized as follows. Section \ref{sec:background} presents background material related to CLFs and CBFs, before Section \ref{sec:qp1} introduces the proposed task-priority framework. Section \ref{sec:simulations} presents simulation results of the framework implemented on an articulated intervention autonomous underwater vehicle (AIAUV), while conclusions and future work can be found in Section \ref{sec:conclusion}.

\section{Background Material}\label{sec:background}
In this section, the necessary background material will be presented. For compactness, we will slightly abuse notation and denote 
\begin{align}
    L_g h(x) = \pdv{h(x)}{x}g(x), \label{eq:notationabuse}
\end{align}
whenever $h(x)$ is a scalar or vector-valued function, and $g(x)$ is a vector field or a matrix. Note that \eqref{eq:notationabuse} is only equal to the Lie derivative of $h(x)$ along $g(x)$ when $h(x)$ is a multivariable scalar function and $g(x)$ a vector field.

\subsection{Model}\label{ssec:model}
Consider the nonlinear affine control system 
\begin{align}
    \dot{x} = f(x) + g(x)u\label{eq:nonlinsys},
\end{align}
where $f$ and $g$ are locally Lipschitz, $x\in D\subset\mathbb{R}^l$ and $u\in U\subset\mathbb{R}^p$ is the set admissible control inputs. Let the locally Lipschitz vector-valued function $y=\sigma(x)-\sigma_d$ describe the error coordinates of some equality task $\sigma:\mathbb{R}^l\to \mathbb{R}^m$. Under the following assumption
\begin{align}
    L_g L_f^k y &= 0,\quad 0\leq k\leq \rho-2\\
    L_g L_f^{\rho-1}y &\neq 0,
\end{align}
the input-output dynamics becomes 
\begin{align}
    y^{(\rho)}= \underbrace{L_f^\rho y(x)}_{b(x)} + \underbrace{L_g L_f^{\rho-1} y(x)}_{A(x)} u . \label{eq:io-dynamics}
\end{align}
The system \eqref{eq:nonlinsys} can be decomposed into transverse dynamics states $\eta = \col \left(y,\dot{y},\dotsc, y^{(\rho-1)} \right)\in X\subset \mathbb{R}^{\rho m}$ and internal dynamics states $z\in Z\subset \mathbb{R}^{l-\rho m}$ as follows
    \begin{subequations}\label{eq:transzero}
        \begin{align}
    % \dot{\eta} &= F\eta+G \left(A(x)u+b(x) \right),\label{eq:taskdyn}\\
    \dot{\eta} &= \bar{f}(\eta,z)+\bar{g}(\eta,z)u,\label{eq:taskdyn}\\
    \dot{z} &= f_z(\eta,z),
        \end{align}
    \end{subequations}
with $\bar{f}(\eta,z) = F\eta +Gb(x)$ and $\bar{g}(\eta,z) = GA(x)$ where 
\begin{align}
    F= \begin{bmatrix}
        0 & I & 0 & \cdots & 0\\
        0 & 0 & I & \cdots & 0\\
        \vdots & \ddots & \ddots & \ddots & \vdots\\
        0 & 0 & 0 & \cdots & I\\
        0 & 0 & 0 & 0 & 0
    \end{bmatrix},\quad G = \begin{bmatrix}
        0\\
        0\\
        0\\
        \vdots\\
        I
    \end{bmatrix},\label{eq:F-G-matrices}
\end{align}
where $0$ is the $m\times m$ matrix of zeros and $I$ is the $m\times m$ identity matrix.
\subsection{Control Lyapunov Functions}
% Since the dimensions of $u$ and $y$ are not necessarily equal, $A(x)$ may not be invertible, and a feedback linearizing control law of the form 
% \begin{align}
%     u &= A^{-1} \left(\mu -b \right),
% \end{align}
% cannot be employed. However, a method based on control Lyapunov functions \citep{AmesRES} can be used for controlling the transverse variables whenever $A(x)$ has full rank.
A control Lyapunov function is a candidate Lyapunov function $V$, for which $\dot{V}$ can be made negative by appropriate selection of the control input $u$. In order to explicitly control the rate of exponential convergence, a specific type of CLF is defined in \cite{AmesRES} as follows:
\begin{defn}
    A continuously differentiable and positive definite function $V_\epsilon : X \to \mathbb{R}$ is said to be a rapidly exponentially stabilizing control Lyapunov function (RES-CLF) for the system \eqref{eq:transzero} if there exists constants $c_1,c_2,c_3 > 0$ such that for all $0<\epsilon < 1$ and for all states $(\eta,z)\in {X\times Z}$ it holds that
    \vspace{-.5em}
    \begin{align}
        c_1\norm{\eta}^2 \leq V_\epsilon (\eta) \leq \frac{c_2}{\epsilon^2}\norm{\eta}^2,\\
        \inf_{u\in U}\left[L_{\bar{f}} V_\epsilon(\eta,z)+L_{\bar{g}}V_\epsilon (\eta,z) u+ \frac{c_3}{\epsilon}V_\epsilon(\eta) \right]\leq 0. \label{eq:res-clf}
    \end{align}
    \vspace{-1.5em}
\end{defn}
Such a function can be constructed by solving the continuous time algebraic Riccati equation
\begin{align}
    F^T P+ PF-PGG^T P + Q = 0,
\end{align}
for $P=P^T > 0$, where $Q$ is any positive definite matrix. In order to stabilize the transverse dynamics at a rate $\epsilon$ define
\begin{align}
    V_\epsilon(\eta) = \eta^T \begin{bmatrix}
        \frac{1}{\epsilon}I & 0\\
        0 & I
    \end{bmatrix}P\begin{bmatrix}
        \frac{1}{\epsilon}I & 0 \\
        0 & I
    \end{bmatrix}\eta:= \eta^T P_\epsilon \eta.\label{eq:clf}
\end{align}
When $A(x)$ has linearly independent rows, it can be shown that the time derivative of \eqref{eq:clf} satisfies \citep{AmesRES}
\begin{align}
	% \inf_{u \in U}\left[L_{\bar{f}} V_{\epsilon}(\eta,z)+L_{\bar{g}} V_{\epsilon}(\eta,z)u+\frac{\gamma}{\epsilon} V_{\epsilon}(\eta) \right] \leq 0,
	\inf_{u \in U}\left[L_{\bar{f}} V_{\epsilon}(\eta,z)+L_{\bar{g}} V_{\epsilon}(\eta,z)u \right] \leq -\frac{\gamma}{\epsilon} V_{\epsilon}(\eta),
\end{align}
where $\gamma := \frac{\lambda_{\text{min}}\left(Q \right)}{\lambda_{\text{max}}\left(P\right) }>0$ and 
\begin{align}
    L_{\bar{f}}V_\epsilon (\eta,z) &= \eta^T \left(F^T P_\epsilon + P_\epsilon F \right)\eta+2\eta^T PGb,\\
    L_{\bar{g}}V_\epsilon(\eta,z)&= 2\eta^T P_\epsilon GA.
\end{align}

\subsection{Control Barrier Functions}
Control objectives described by inequalities or sets can be enforced by rendering the superlevel set
\begin{align}
    \mathcal{C} = \left\{x\in D\subset \mathbb{R}^l : h(x) \geq 0 \right\},
\end{align}
of some continuously differentiable function $h:D\to \mathbb{R}$ forward invariant \citep{Ames2019}.
\begin{defn}\label{def:cbf}
    Let $\mathcal{C}\subset D\subset\mathbb{R}^l$ be the superlevel set of a continuously differentiable function $h : D\to \mathbb{R}$, then $h$ is a control barrier function (CBF) for the system \eqref{eq:nonlinsys} if there exists an extended class $\mathcal{K}_\infty$ function $\alpha$ such that 
    \begin{align}
        \sup_{u\in U}\left[L_f h(x) + L_g h(x) u \right]\geq -\alpha\left(h(x)\right),
    \end{align}
    for all $x\in D$.
\end{defn}
The existence of a CBF implies that the superlevel set $\mathcal{C}$ of the function $h$ can be rendered forward invariant by appropriate selection of the control input \citep{Ames2017}, which means that if $x(t_0)=x_0\in \mathcal{C}$, then $x=x(t)\in \mathcal{C} $ for all $t\geq t_0$. Equivalently, if $h(x_0)\geq 0$, then $h(x) \geq 0$ for all $t\geq t_0$.

\subsection{Exponential Control Barrier Functions}
Definition \ref{def:cbf} assumes that the relative degree of $h$ is equal to one. However, safety-related tasks for robotic systems are often a function of the configuration variables only, meaning that they have a higher relative degree. Introduced in \cite{Nguyen2016} and refined in \cite{Ames2019}, exponential control barrier functions generalizes CBFs to functions $h(x)$ with arbitrary relative degree $r\geq 1$. To this end, we define $\eta_b = \col \left(h(x),L_f h(x), L_f^2h(x),\dotsc, L_f^{r-1}h(x) \right)$ and assume that $u$ can be chosen such that $L_f^r h(x) +L_gL_f^{r-1}h(x)u=\mu$ for a given $\mu \in U_\mu\subset \mathbb{R}$. The dynamics of $h(x)$ can then be written as a linear system 
\begin{align}
    \dot{\eta}_b(x) &= F\eta_b(x)+G\mu,\\
    h(x) &= C\eta_b(x),
\end{align}
where
\begin{align}
    F_b &= \begin{bmatrix}
        0 & 1 & 0 & \cdots & 0\\
        0 & 0 & 1 & \cdots & 0\\
        \vdots & \vdots & \vdots & \ddots & \vdots\\
        0 & 0 & 0 & \cdots & 1\\
        0 & 0 & 0 & \cdots & 0
    \end{bmatrix},\quad G_b= \begin{bmatrix}
        0\\
        0\\
        \vdots\\
        0\\
        1
    \end{bmatrix},\\
    C_b &= \begin{bmatrix}
        1 & 0 & \cdots & 0
    \end{bmatrix}.
\end{align}
Choosing the state feedback $\mu=-K_\alpha \eta_b(x)$ results in $h(x(t)) = C_be^{(F_b-G_bK_\alpha)t}\eta_b(x_0)$, which by the comparison lemma implies that if $\mu \geq -K_\alpha \eta_b(x)$ then $h(x(t))\geq C_be^{(F_b-G_bK_\alpha)t}\eta_b(x_0)$. 
\begin{defn}\label{def:ecbf}
    Given a set $\mathcal{C}\subset D\subset \mathbb{R}^l$ defined as the superlevel set of an $r$-times continuously differentiable function $h:D\to\mathbb{R}$, then $h$ is an exponential control barrier function (ECBF) for the control system \eqref{eq:nonlinsys} if there exists a row vector $K_\alpha \in\mathbb{R}^r$ such that
    \begin{align}
        \sup_{u\in U}\left[L_f^rh(x) + L_g L_f^{r-1}h(x)u \right]\geq -K_\alpha \eta_b (x), \label{eq:ecbf}
    \end{align}
    $\forall\, x\in \Int \left(\mathcal{C}\right)$ results in $h(x(t))\geq C_be^{(F_b-G_bK_\alpha)t}\eta_b(x_0)\geq 0$ whenever $h(x_0)\geq 0$.
\end{defn}

\subsection{Combining CLFs and ECBFs}
The RES-CLF and ECBF conditions in \eqref{eq:res-clf} and \eqref{eq:ecbf} are both affine in the control input $u$, which means that the control problem can be formulated as a convex optimization problem, enabling the incorporation of control input saturation limits and rate constraints \citep{Galloway2015}. By employing RES-CLFs, the CLF-ECBF-based QP from \citep{Ames2017,Ames2019} becomes:
\begin{mini}[3]
    {\substack{u\in\mathbb{R}^m,\delta\in\mathbb{R}}}{\frac{1}{2}u^T H(x) u + c^T(x) u+ w\delta^2}{\label{eq:res-clf-ecbf-qp}}{}
    \addConstraint{L_{\bar{f}} V_\epsilon(\eta,z)+L_{\bar{g}}V_\epsilon (\eta,z) u }{\leq -\frac{\gamma}{\epsilon}V_{\epsilon}+\delta}
    \addConstraint{L_f^r h(x) + L_g L_f^{r-1}h(x)u}{\geq -K_\alpha \eta_b (x),}
\end{mini}  
where $H\colon D\to \mathbb{R}^{m\times m}$ is any positive semi-definite matrix, $c\colon D\to \mathbb{R}^m$, and $\delta\in\mathbb{R}$ is a slack variable penalized by $w>0$, ensuring the feasibility of the QP in case of conflicting set-based and equality-based control objectives.

\section{Quadratic Programs for $N$ Equality- and $M$ Set-Based Control Tasks} \label{sec:qp1}
This section extends the CLF-ECBF QP controller in \eqref{eq:res-clf-ecbf-qp} to an arbitrary number of equality- and set-based control tasks distributed to an arbitrary number of priority levels.
\subsection{CLF Penalty Parameters as a Priority Measure}\label{ssec:ext1}
% \todo[inline]{Can we remove eq (19)? why is it needed? Replacing (19) by (23) directly?}
% Inspired by \citep{Ames2013}, the QP in \eqref{eq:res-clf-ecbf-qp} can be extended to $N$ equality-based control objectives by deriving the input-output dynamics for each control objective as in Section \ref{ssec:model} and defining 
% \begin{align}
%     \begin{bmatrix}
%         y_1^{(\rho_1)}\\
%         y_2^{(\rho_2)}\\
%         \vdots\\
%         y_N^{(\rho_N)}
%     \end{bmatrix}=\underbrace{\begin{bmatrix}
%         L_f^{\rho_1} y_1(x)\\
%         L_f^{\rho_2} y_2(x)\\
%         \vdots\\
%         L_f^{\rho_N} y_N(x) 
%     \end{bmatrix}}_{b(x)} + \underbrace{\begin{bmatrix}
%         L_g L_f^{\rho_1-1} y_1(x)\\
%         L_g L_f^{\rho_2-1} y_2(x)\\
%         \vdots\\
%         L_g L_f^{\rho_N-1}y_N(x)
%     \end{bmatrix}
%         }_{A(x)} u.\label{eq:N-task-dyn}
% \end{align}
Inspired by \cite{Ames2013}, the QP in \eqref{eq:res-clf-ecbf-qp} can be extended to $N$ equality-based control objectives by deriving the input-output dynamics for each control objective, i.e.
\begin{align}
    y_i^{(\rho_i)}(x) =\underbrace{L_f^{\rho_i}y_i(x)}_{b_i(x)} +\underbrace{L_gL_f^{\rho_i-1}y_i(x)}_{A_i(x)}u, \label{eq:io-dyn}
\end{align}
for each $i=1,\dotsc,N$. Transverse dynamics states $\eta_i = \col \big(y_i, \dot{y}_i,\dotsc, {y}_i^{(\rho_i-1)} \big)$ and RES-CLFs $V_{\epsilon,i}$ can then be defined analogously to \eqref{eq:taskdyn}, \eqref{eq:F-G-matrices} and \eqref{eq:clf}.
Moreover, $M$ set-based tasks described by the superlevel set $\mathcal{C}_j$ of some $r_j$ times continuously differentiable function $h_j(x)$ can be included at the highest priority level (which is implied by no slack variables). The control input can then be obtained from the QP:
\begin{mini}[3]
    % {\substack{u\in\mathbb{R}^m,\delta\in\mathbb{R}^N}}{u^T A^T A u + 2b^T A u +\delta^T W \delta }{\label{eq:taskpri-ecbf-clf-qp}}{}
    {\substack{\left(u,\delta\right)\in\mathbb{R}^{m+N}}}{u^T H(x) u + c^T(x) u+\delta^T W \delta }{\label{eq:taskpri-ecbf-clf-qp}}{}
    \addConstraint{L_{\bar{f}_i} V_{\epsilon,i}+L_{\bar{g}_i}V_{\epsilon,i}  u }{\leq -\frac{\gamma_i}{\epsilon}V_{\epsilon,i}+\delta_i ,}{i=1,\ldots,N}
  \addConstraint{L_f^{r_k} h_k + L_g L_f^{r_k-1}h_k u}{\geq -K_{\alpha,k} \eta_{b,k},\quad}{k=1,\dotsc,M,}
\end{mini}
where $W\in\mathbb{R}^{N\times N}$ is a diagonal matrix of penalty parameters, $\eta_{b,k}\hspace{-.5mm}=\hspace{-0.5mm}\col \big(h_k(x),L_fh_k(x),\dots,L_f^{r-1}h_k(x) \big)$ and
\begin{align}
    L_{\bar{f}_i}V_{\epsilon,i} &= \eta_i^T \left(F_i^T P_{\epsilon,i} + P_{\epsilon,i} F_i \right)\eta_i+2\eta_i^T P_iG_ib_i,\\
    L_{\bar{g}_i}V_{\epsilon,i}&= 2\eta_i^T P_{\epsilon,i} G_iA_i.
\end{align}

The equality tasks encoded by RES-CLFs are prioritized by adjusting the elements of the diagonal penalty matrix $W$. The satisfaction of all equality tasks are therefore described by a single objective function through the value of the slack variables $\delta$ and the penalty parameters in $W$. Whenever equality tasks are incompatible, this fact invariably leads to trade-off configurations that do not satisfy any of the tasks. Hence, strict priority between tasks cannot be achieved in the sense that lower-priority tasks have no effect on the execution of higher-priority tasks. As a result, it is challenging to include lower-priority optimization-based tasks since they will interfere with more critical higher-priority tasks such as end-effector control whenever the tasks are incompatible. 
\subsection{Main Result: Enforcing Strict Priority Between a Selection of Tasks}
In order to establish more than two strict priority levels, we propose to solve a quadratic program for every priority level as suggested for kinematic control in \cite{Kanoun2011}. The idea is to begin by computing a control input according to \eqref{eq:taskpri-ecbf-clf-qp} that only accounts for safety-related set-based tasks and equality tasks at the highest priority level. Subsequently, a new quadratic program is solved for each priority level, refining the previous solution in an attempt to satisfy lower-priority tasks without affecting the execution of higher-priority tasks. 

Consider $N$ equality tasks and $M$ set-based tasks distributed to $k$ priority levels, with $N=N_1+\dotsc+N_k$ and $M=M_1+\dotsc + M_k$, where $N_i$ and $M_i$ denotes the number of equality and set-based tasks at priority level $i$, respectively. A control input $u_1^*$ that disregards all lower-priority tasks is obtained by solving \eqref{eq:taskpri-ecbf-clf-qp} with $i=1,\dotsc,N_1$ and $k=1,\dotsc,M_1$. If the system is redundant with respect to these $N_1+M_1$ tasks, the control input $u_1^*$ can be refined without affecting how the $N_1$ higher-priority equality tasks are executed by enforcing 
\begin{align}
    L_{\bar{f}_i} V_{\epsilon,i}+L_{\bar{g}_i}V_{\epsilon,i}  u &\leq L_{\bar{f}_i} V_{\epsilon,i}+L_{\bar{g}_i}V_{\epsilon,i}  u_1^*
\end{align}
which implies that $L_{\bar{g}_i}V_{\epsilon,i}u \leq L_{\bar{g}_i}V_{\epsilon,i}u_1^*$ for all $i=1,\dotsc,N_1$. Similarly, the higher-priority set-based tasks are unaffected by enforcing 
\begin{align}
    L_g L_f^{r_k-1}h_k u \geq L_gL_f^{r_k-1}h_k u_{1}^*,
\end{align}
for all $k=1,\dotsc,M_1$. Consider $N_2$ additional equality-based tasks and $M_2$ additional set-based tasks. The control input $u_1^*$ can be modified to account for lower-priority tasks without affecting how the $N_1$ and $M_1$ equality- and set-based tasks are executed by solving:
\begin{mini}[3]
    {\scriptstyle{(u,\delta,s)\in\mathbb{R}^{m+N_2+M_2} }}{u^T H(x) u + c^T(x) u +\delta^T W_2 \delta+s^T K_2 s }{\label{eq:qp-2}}{}
    \addConstraint{L_{\bar{g}_i}V_{\epsilon,i}u }{\leq L_{\bar{g}_i}V_{\epsilon,i}u_1^*, }{\scriptstyle{i=1,\dotsc,N_1}}
    \addConstraint{L_{\bar{f}_j} V_{\epsilon,j}+L_{\bar{g}_j}V_{\epsilon,j}  u }{\leq -\frac{\gamma_j}{\epsilon}V_{\epsilon,j}+\delta_j ,}{\scriptstyle{j=N_1+1,\ldots,N_1+N_2}}
    % \addConstraint{L_f^{r_k} h_k + L_g L_f^{r_k-1}h_k u}{\geq -K_{\alpha,k} \eta_{b,k},}{\scriptstyle{k=1,\dotsc,M_1}}
    \addConstraint{L_g L_f^{r_k-1}h_k u}{\geq L_gL_f^{r_k-1}h_k u_{1}^*,}{\scriptstyle{k=1,\dotsc,M_1}}
    \addConstraint{L_f^{r_l} h_l + L_g L_f^{r_l-1}h_l u}{\geq -K_{\alpha,l} \eta_{b,l} - s_l,\quad}{\scriptstyle{l=M_1+1,\dotsc,M_1+M_2,}} 
\end{mini}
where slack variables $s$ penalized by the elements in the diagonal matrix $K>0$ have been added to the lower-priority set-based tasks enforced through ECBFs to ensure feasibility of the optimization problem. 

By observing that the solution $u_2^*$ to \eqref{eq:qp-2} enforces the constraints $L_{\bar{g}_i}V_{\epsilon,i}u_2^*\leq L_{\bar{g}_i}V_{\epsilon,i}u_1^*$ and $L_g L_f^{r_k-1}h_k u_2^* \geq L_g L_f^{r_k-1}h_k u_1^*$ for all $i$ and $k$, it is straightforward to generalize \eqref{eq:qp-2} to an arbitrary priority level $n$:
\begin{mini}[3]
    {\substack{\left(u,\delta, s\right)\in\mathbb{R}^{m+N_n+M_n} }}{u^T H u + c^T u +\delta^T W_n \delta+s^T K_n s }{\label{eq:qp-n}}{}
    \addConstraint{L_{\bar{g}_i}V_{\epsilon,i}u }{\leq L_{\bar{g}_i}V_{\epsilon,i}u_{n-1}^*, }{\scriptstyle{i=1,\dotsc,\bar{N}_{n-1}}}
    \addConstraint{L_{\bar{f}_j} V_{\epsilon,j}+L_{\bar{g}_j}V_{\epsilon,j}  u }{\leq -\frac{\gamma_j}{\epsilon}V_{\epsilon,j}+\delta_j ,}{\scriptstyle{j=\bar{N}_{n-1}+1,\ldots,\bar{N}_n}}
    % \addConstraint{L_f^{r_k} h_k + L_g L_f^{r_k-1}h_k u}{\geq -K_{\alpha,k} \eta_{b,k},}{\scriptstyle{k=1,\dotsc,M_1}}
    \addConstraint{L_g L_f^{r_k-1}h_k u}{\geq L_g L_f^{r_k-1}h_k u_{n-1}^* ,\quad}{\scriptstyle{k=1,\dotsc,\bar{M}_{n-1} }}
    \addConstraint{L_f^{r_l} h_l + L_g L_f^{r_l-1}h_l u}{\geq -K_{\alpha,l} \eta_{b,l} - s_l,\quad}{\scriptstyle{l=\bar{M}_{n-1}+1,\dotsc,\bar{M}_n,}}
\end{mini}
where $\bar{N}_n = N_1+N_2+\ldots + N_{n}$ and $\bar{M}_n=M_1+M_2+\ldots + M_{n}$. Note that the objective function is slightly different at every priority level, since the slack variables $\delta$ and $s$ always correspond to tasks at the current priority level. This prevents trade-off configurations where none of the tasks are satisfied from occurring. The procedure is summarized in Algorithm \ref{alg:hierarchicaltask}.

\begin{algorithm}
	\caption{Task priority CLF-ECBF QP controller}
	\begin{algorithmic}[1]
		\renewcommand{\algorithmicrequire}{\small \textbf{Input:}}
		\renewcommand{\algorithmicensure}{\small \textbf{Output:}}
        % \REQUIRE $H(x)$, $c(x)$, $N$ equality tasks described by CLFs $V_{\epsilon,i}$ and $M$ set-based tasks described by ECBFs $h_i(x)$.
        \Require \small $H(x)$, $c(x)$, $V_{\epsilon,i}(\eta_i)$, $i=1,\dotsc,N$, $h_j(x)$, $j=1,\dotsc,M$.
        \Ensure \small $u$
        \State Solve \eqref{eq:taskpri-ecbf-clf-qp} to obtain $u_1^*$ with $i=1,\dotsc,N_1$, $k=1,\dotsc,M_1$.
        \For {$p=2$ to $k$}
        \State Solve \eqref{eq:qp-n} to obtain $u_p^*$.
        \EndFor
        \State \textbf{return} $u=u_k^*$.
	\end{algorithmic}\label{alg:hierarchicaltask}
\end{algorithm}
\vspace{-.5em}
\section{Simulations}\label{sec:simulations}
\begin{figure}
    \centering
    \includegraphics[width=0.3\textwidth]{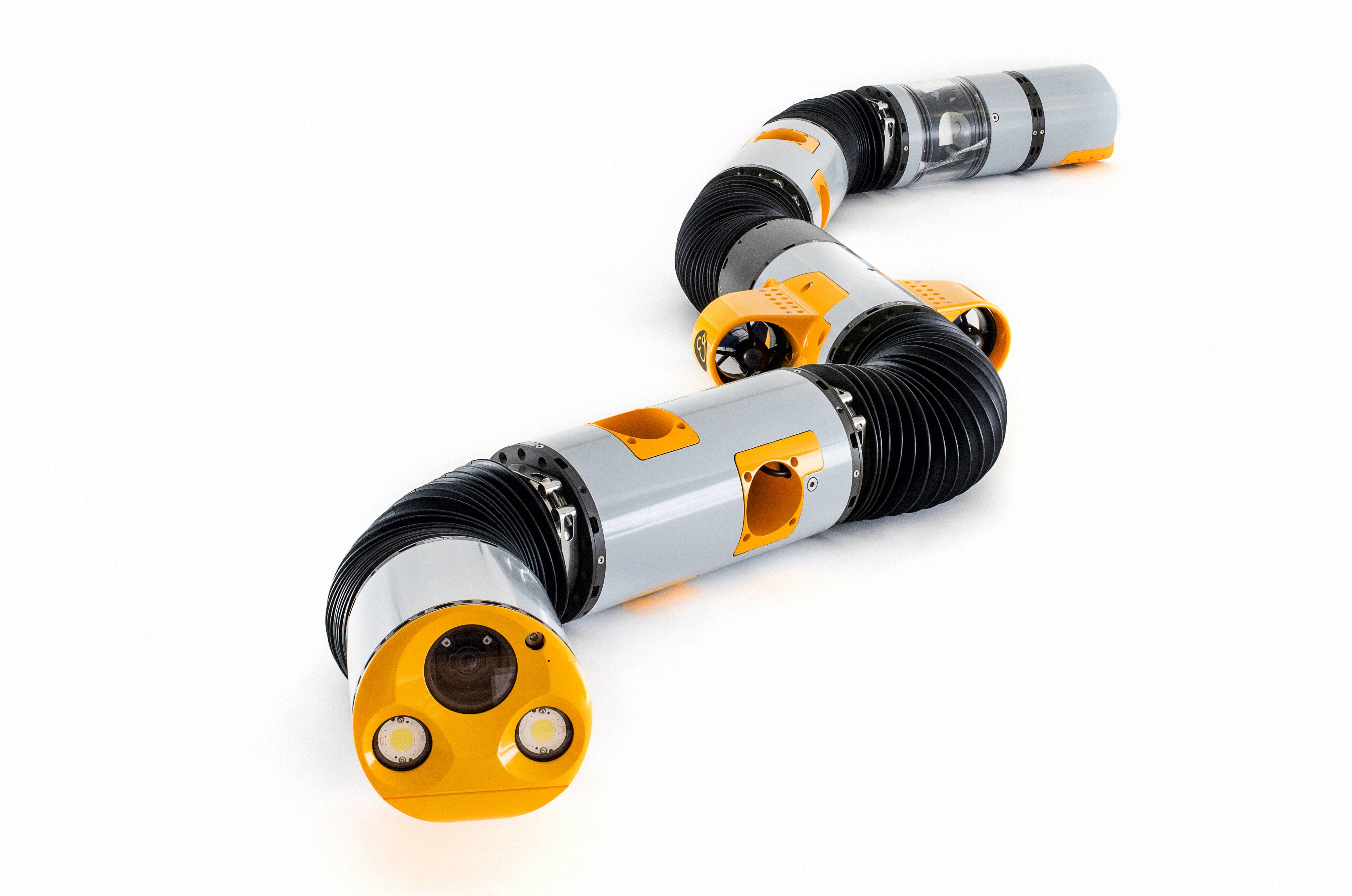}
    \vspace{-1em}
    \caption{The Eelume AIAUV (Courtesy of Eelume)}
    \label{fig:eelume}
\end{figure}

\vspace{-0.3em}
% \todo[inline]{Links numbered ... from tail to head...}
In this section, the proposed hierarchical control scheme is validated in simulation on an AIAUV based on the Eelume robot \citep{Schmidt-Didlaukies2018,liljeback17} depicted in Figure \ref{fig:eelume}. The AIAUV is a floating base manipulator, with $n+1$ links interconnected by $n$ joints, where link $1$ is the tail, or base link and link $n+1$ is the head. The simulation model has $n=8$ single DOF and revolute joints and $p=7$ thrusters. The system configuration is described by $\xi= \col\left(p_{ib}^i,q,\theta \right)\in\mathbb{R}^{7+n}$, where $p_{ib}^i\in\mathbb{R}^3$ is the position of the base of the AIAUV in an inertial frame, $q=\col \left(\eta,\epsilon\right)\in\mathbb{R}^4$ is a unit quaternion describing the orientation of the base and $\theta=\col \left(\theta_1,\dots,\theta_n\right)\in\mathbb{R}^n$ are the joint angles. The joint velocities are given by $\dot{\theta}$ and the linear and angular velocities of the base frame with respect to an inertial frame are denoted $v_{ib}^b$ and $\omega_{ib}^b$, respectively. These velocities are collected in the velocity vector $\zeta=\col\big(
    v_{ib}^b,
    \omega_{ib}^b,\dot{\theta}\big)\in\mathbb{R}^{6+n}$. The equations of motion are given by \citep{Schmidt-Didlaukies2018}
\begin{align}
        \dot{\xi}= J_\xi (q) \zeta,\\
        M(\theta) \dot{\zeta}+C(\theta,\zeta)\zeta + D(\theta,\zeta)\zeta + g(\xi) = B(\theta)u,
\end{align}
where $M(\theta)$ is the inertia matrix including hydrodynamic added mass, $C(\theta,\zeta)$ is the Coriolis-centripetal matrix including hydrodynamic added mass, $D(\theta,\zeta)$ is the damping matrix, $g(\xi)$ is the vector of gravitational and buoyancy forces and moments, $B(\theta)$ is the actuator configuration matrix and $u=\col\left(u_t,u_j \right)\in\mathbb{R}^{p+n}$ consists of the thruster inputs $u_t\in\mathbb{R}^p$ and joint torque inputs $u_j\in\mathbb{R}^n$. Moreover, the kinematic transformation matrix is given by  
\begin{align}
        \hspace{-0.8em}J_\xi(q) = \begin{bmatrix}
            R^i_b(q) & 0_{3\times 3} & 0_{3\times n}\\
            0_{4\times 3} & T_q(q) & 0_{4\times n}\\
            0_{n\times 3} & 0_{n\times 3} & I_n
        \end{bmatrix}\hspace{-0.25em}, T_q(q) = \frac{1}{2}\begin{bmatrix}
            -\epsilon^T\\
            \eta I_3 + \left[\epsilon \right]_\times
        \end{bmatrix},
\end{align}
where $R^i_b(q)\in \specialo(3)$ is a rotation matrix describing the rotation between the base and inertial frame and $\left[\cdot \right]_\times :\mathbb{R}^3\to\mathfrak{so}(3)\subset \mathbb{R}^{3\times 3}$ denotes the skew symmetric map. 
    
By defining $x=\col\left(x_1,x_2 \right)=\col\left(\xi,\zeta \right)$, the equations of motion can be rewritten in state space form 
\begin{align}
    \dot{x} = f(x) + g(x)u,
\end{align}
where 
\begin{align}
    f(x) &= \begin{bmatrix}
        J_\xi (x_1)x_2\\
        -M(x_1)^{-1} \left(C(x)x_2+D(x)x_2+g(x_1) \right)
    \end{bmatrix},\\
    g(x) &= \begin{bmatrix}
        0\\
        M(x_1)^{-1}B(x_1)
    \end{bmatrix}.
\end{align}
With $6+n=14$ DOFs and $p+n=15$ control inputs, the system is overactuated, since the number of actuators is greater than the number of DOFs. Moreover, $14$ DOFs imply that the system is redundant with respect to typical tasks such as end-effector configuration control.

% For an AIAUV, operational space tasks are defined by $\sigma=f_\sigma(\xi)$, and the Jacobians by 
% \begin{align}
%     \dot{\sigma}(\xi,\zeta) &= \pdv{f_\sigma(\xi)}{\xi}J_\xi (q)\zeta\\
%     &= J \zeta.
% \end{align}
We consider four equality-based tasks and three set-based tasks, at three different priority levels. The set-based tasks are safety-related and are thus placed at the highest priority level. The safety-related tasks consist of end-effector collision avoidance, actuator configuration matrix singularity avoidance and joint limit avoidance. The purpose of the end-effector collision avoidance task is to avoid a spherical obstacle with radius $r_{\text{obs}}\in\mathbb{R}$. To this end, the scalar distance measure between the center of the obstacle and the end-effector is employed as a set-based task $\sigma_a\in\mathbb{R}$. In order to ensure that the distance from the end-effector to the center of the spherical obstacle is always greater than some lower limit, we enforce the positivity of the following ECBF
\begin{align}
    h_1 &= \underbrace{\sqrt{\left(p_{\text{obs}}^i-p_{ie}^i \right)^T \left(p_{\text{obs}}^i-p_{ie}^i \right)}}_{\sigma_a} - (r_{\text{obs}}+\epsilon),\label{eq:ecbf1}
\end{align}
where $\epsilon\in\mathbb{R}$ defines an inaccessible safety region around the spherical obstacle and $p_{ie}^i$ and $p_{\text{obs}}^i$ are the positions of the end-effector and the center of the spherical obstacle in an inertial frame, respectively.
%and are all encoded by ECBFs with a relative degree of two, since they are only functions of the configuration variables.

Rank deficiency of the actuator configuration matrix $B(\theta)$ was pointed out in \cite{SverdKelas_18}, and implies that no force or moment can be generated in certain directions in the vector space $\mathbb{R}^{6+n}$ belonging to $\tau$. Inspired by the manipulability measure \citep{yoshikawa85}, the actuation measure $\sigma_b=\det\left(B(\theta)B^T(\theta) \right)$ is introduced as a high-priority set-based task to prevent singular configurations. The actuation measure is kept above a minimum value $\sigma_{b,\text{min}}$ through the following ECBF
\begin{align}
    h_{2} &= {\det \left(B(\theta)B^T(\theta) \right)} - \sigma_{b,\text{min}}.
\end{align}
The third safety-related task is the joint limit avoidance task $\sigma_c=\theta\in\mathbb{R}^n$, which has both lower and upper limits. Hence, $2n$ ECBFs are needed of the form
\begin{align}
    h_{j+2} &= \theta_j - \theta_{j,\text{min}},\\
    h_{j+2+n} &= \theta_{j,\text{max}}-\theta_j.\label{eq:ecbf4}
\end{align}
for $j=1,\dots,n$. 

The second priority level contains the equality-based end-effector positioning and orientation tasks
\begin{align}
    y_1 &= p_{ie}^i-p^i_{d,e},\\
    y_2 &= \tilde{\epsilon},
\end{align}
where $p_{d,e}^i$ is the desired end-effector position and $\tilde{\epsilon}$ is the imaginary part of the quaternion error vector $\tilde{q}=q_d\otimes q^*$, which is given by 
\begin{align}
    \tilde{\epsilon} = \eta \epsilon_d -\eta_d \epsilon + \left[\epsilon\right]_\times \epsilon_d,
\end{align}
where $q_d$ and $q$ are the quaternion representations of the desired and measured orientation of the end-effector, respectively. 
% Moreover, $\left[\cdot \right]_\times :\mathbb{R}^3\to\mathfrak{so}(3)\subset \mathbb{R}^{3\times 3}$ denotes the skew symmetric map.

In order to minimize movement of the base while repositioning the end-effector, a base positioning task is defined at the third priority level
% At the third and final priority level we find a base positioning task, intended to keep the base of the AIAUV stationary while repositioning the end-effector. 
\begin{align}
    y_3 &= p^i_{ib}-p^i_{d,b},
\end{align}
where $p_{ib}^i\in\mathbb{R}^3$ and $p_{d,b}^i\in\mathbb{R}^3$ are the measured and desired positions of the AIAUV base in the inertial frame, respectively. 
Note that the end-effector positioning and orientation and base positioning tasks only consume 9 DOFs, which entails that there are still 5 uncontrolled DOFs if all set-based tasks are inactive. Stability of the entire system can therefore only be guaranteed if the resulting zero dynamics is asymptotically stable. Instead of performing a complicated analysis of the zero dynamics, a joint velocity regulation task is designed to eliminate the residual DOFs of the system $y_4 = \dot{\theta}$, where $\dot{\theta}\in\mathbb{R}^n$ is the vector of joint velocities.
% Because $y_1,y_2$ and $y_3$ only depend on the configuration variables $\xi$, they have to be differentiated twice with respect to time for the input $u$ to appear. The joint velocity regulation task $y_4$ is a function of the generalized velocity $\zeta$, and thus needs to be differentiated once for the input to show up. Hence, $\rho_1=\rho_2=\rho_3=2$ and the input-output dynamics can be obtained from \eqref{eq:io-dyn}, transverse dynamics states $\eta_i=\col \left(y_i,\dots,y_i^{(\rho_i-1)} \right)$ and RES-CLFs $V_{\epsilon,i}$ can be defined analogously to \eqref{eq:taskdyn}, \eqref{eq:F-G-matrices} and \eqref{eq:clf}  and $\rho_4=1$. Similarly, \eqref{eq:ecbf1}-\eqref{eq:ecbf4} all have to be differentiated twice with respect to time for the input to show up, hence $r_1=\dots=r_{18}=2$. 

The input-output dynamics of the equality tasks are then obtained from \eqref{eq:io-dyn} such that transverse dynamics states $\eta_i$ and RES-CLFs $V_{\epsilon,i}$ can be defined analogously to \eqref{eq:taskdyn}, \eqref{eq:F-G-matrices} and \eqref{eq:clf}, with $\rho_1=\rho_2=\rho_3=2$ and $\rho_4=1$, where $\rho_i$ denotes the amount of times $y_i$ has to be differentiated for the input to appear. Furthermore, the set-based tasks in \eqref{eq:ecbf1}-\eqref{eq:ecbf4} all have to be differentiated twice with respect to time for the input to show up, hence $r_1=\dots=r_{18}=2$. 

The design matrix $H(x)$ and design vector $c(x)$ in the objective functions are selected by minimizing the virtual control input $\mu = Au+b$ quadratically as done in \cite{Ames2013,Galloway2015}, where 
\begin{align}
    A(x) = \begin{bmatrix}
        A_1(x)\\
        A_2(x)\\
        A_3(x)\\
        A_4(x)
    \end{bmatrix},\quad b(x) = \begin{bmatrix}
        b_1(x)\\
        b_2(x)\\
        b_3(x)\\
        b_4(x)
    \end{bmatrix}.
\end{align}
In terms of $u$, this yields 
\begin{align}
    \mu^T\mu = u^TA^T Au+ 2b^T Au+ b^Tb,
\end{align}
which implies that $H(x)=A^T(x)A(x)$ and $c^T(x) = 2b^T(x)A(x)$. 

According to Algorithm \ref{alg:hierarchicaltask}, we solve the following QP:
\begin{mini}[3]
    % {\substack{u\in\mathbb{R}^m,\delta\in\mathbb{R}^N}}{u^T A^T A u + 2b^T A u +\delta^T W \delta }{\label{eq:taskpri-ecbf-clf-qp}}{}
    {\substack{u\in\mathbb{R}^{15},\left(\delta_1,\delta_2\right)\in\mathbb{R}^2}}{u^T A^T A u + 2b^T A u+w_1\delta_1^2+w_2\delta_2^2 }{\label{eq:aiauv-qp-1}}{}
    \addConstraint{L_{\bar{f}_i} V_{\epsilon,i}+L_{\bar{g}_i}V_{\epsilon,i}  u }{\leq -\frac{\gamma_1}{\epsilon_i}V_{\epsilon,i}+\delta_i,\quad }{i=1,2}
  \addConstraint{L_f^{2} h_k + L_g L_fh_k u}{\geq -K_{\alpha,k} \eta_{b,k},\quad}{k=1,\dotsc,18}
  \addConstraint{}{-u_{\text{\scaleto{max}{2.5pt}}} \leq u\leq u_{\text{\scaleto{max}{2.5pt}}}}
  \addConstraint{}{-\Delta u_{\text{\scaleto{max}{2.5pt}}}\leq \Delta u\leq \Delta u_{\text{\scaleto{max}{2.5pt}}}}{},
\end{mini}
where $\Delta u= u-u_{\text{prev}}$ is the change in control input, $u_{\text{prev}}$ is the control input at the last sample, and $u_{\text{\scaleto{max}{2.2pt}}}=\col \left(50,\dotsc, 50 \right)$ and $\Delta u_{\text{\scaleto{max}{2.2pt}}}=\col \left(0.1,\dotsc,0.1 \right)$ are thruster and joint torque limits and rate constraints, respectively. The QP in \eqref{eq:aiauv-qp-1} yields a control input $u=u_1^*$ that only accounts for the safety-related tasks and the end-effector positioning and orientation tasks. 
% However, since the end-effector configuration task only consumes 6 DOFs, we may potentially have 8 remaining uncontrolled DOFs. 
The solution $u_1^*$ is refined by utilizing the excess DOFs of the system in an attempt to keep the base stationary and minimize the joint velocities through the QP:
\begin{mini}[3]
    {\scriptstyle{u\in\mathbb{R}^{15},\left(\delta_3,\delta_4\right)\in\mathbb{R}^{2} }}{u^T A^T A u + 2b^T A u +w_3\delta_3^2+w_4\delta_4^2 }{\label{eq:aiauv-qp2}}{}
    \addConstraint{L_{\bar{g}_i}V_{\epsilon,i}u }{\leq L_{\bar{g}_i}V_{\epsilon,i}u_i^*,\quad}{i=1,2}
    \addConstraint{L_{\bar{f}_j} V_{\epsilon,j}+L_{\bar{g}_j}V_{\epsilon,j}  u }{\leq -\frac{\gamma_j}{\epsilon_j}V_{\epsilon,j}+\delta_j,\quad }{{j=3,4}}
    % \addConstraint{L_f^{r_k} h_k + L_g L_f^{r_k-1}h_k u}{\geq -K_{\alpha,k} \eta_{b,k},}{\scriptstyle{k=1,\dotsc,M_1}}
    \addConstraint{L_g L_fh_k u}{\geq L_gL_fh_k u_{1}^*,}{{k=1,\dotsc,18}}
    \addConstraint{}{-u_{\text{\scaleto{max}{2.5pt}}} \leq u\leq u_{\text{\scaleto{max}{2.5pt}}}}
    \addConstraint{}{-\Delta u_{\text{\scaleto{max}{2.5pt}}}\leq \Delta u\leq \Delta u_{\text{\scaleto{max}{2.5pt}}}}{},
\end{mini}
which yields the final control input $u=u_2^*$ that is applied to the AIAUV. The equality task control parameters are listed in Table \ref{tab:control-param-eq}, while $K_{\alpha,k} = \left[3, 4\right]$ for all $k=1,\dots,18$.

\begin{table}
    \centering
    \caption{Equality task convergence rates $\epsilon$ and penalty parameters $w$}
    % \caption{Simulation parameters}
    \begin{tabular}{c|cccc}
        & $y_{1}$ & $y_{2}$ &  $y_{3}$ & $y_{4}$ \\
      \hline
      $\epsilon$ &  $1.2$&$0.2$ & $1.2$ & $0.5$ \\
      $w$ & 60 & 60 & 10 & 10  
    % $\epsilon$ &  $0.78$&$0.1$ & $1.6$ & $1$ \\
    % $w$ & $3$ & $3$ & $3$ & $3$  
    %   \bottomrule
    \end{tabular}
    \label{tab:control-param-eq}
\end{table}

We remark that the optimization problems are formulated in terms of the thruster and joint torque control inputs $u$, and not the commanded forces and torques $\tau=Bu$. Consequently, the proposed framework also solves the control allocation problem, which had to be solved separately in previous works \citep{SverdKelas_18,borlaug2019}. By unifying redundancy resolution, dynamic control and control allocation, strict priority among tasks can always be ensured. The same guarantee does not hold for redundancy resolution schemes that decouple dynamic control and control allocation, since the commanded forces and torques may not be exactly allocable, leading to a loss of priority among tasks.

Simulations were performed in Matlab/Simulink using the ode3 solver with a fixed step-size of 0.01. Simulation results are presented in Figs. 2 to 5. From Figs. 4 and 5 we observe that the high-priority set-based tasks are satisfied at all times. In general, the redundancy of the system is exploited such that the lower-priority equality tasks are satisfied even when higher-priority set-based tasks are at their limits and consuming DOFs. For instance, the actuation measure is kept above a minimum value of $0.1$, which avoids singular configurations of the actuation configuration matrix from occurring and thereby reducing the magnitude and/or rates of change of the control inputs, at the cost of maneuverability \citep{johansen2004}. However, the high-priority collision avoidance task results in a small deviation in the lower-priority end-effector positioning task. Specifically, we observe from Fig. 2 and Fig. 4 that the $x$-coordinate of the end-effector position deviates slightly from its reference and that the distance to the center of the spherical obstacle is at its minimum value between $t\simeq \SI{268}{\second}$ and $t\simeq\SI{281}{\second}$.
% This is explained by referring to \Cref{fig:north-west}, where it is clear that the distance from the end-effector to the spherical obstacle is at its minimum.

After $t\geq \SI{350}{\second}$, the end-effector position is commanded outside of the manipulator workspace (when the base is kept at its current position), which implies that the lower-priority base positioning task is no longer compatible with the higher-priority end-effector positioning task. As desired, the strict priority between tasks is kept at all time, such that the end-effector position converges to its desired value at the expense of a greater error in the base position. 

Finally, we note from Fig. 3 that the thruster and joint torque control inputs are smooth and well within the physical limitations of the Eelume robot.

\begin{figure}
    \centering
    \includegraphics[width=0.33\textwidth]{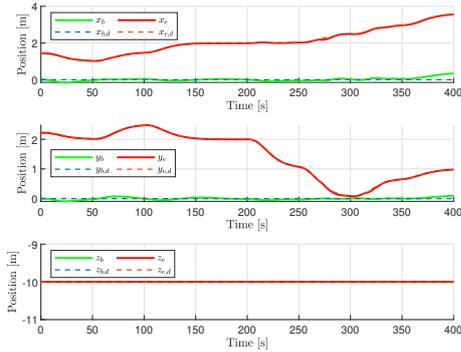}
    \vspace{-1em}
    \caption{The position of the end-effector $p_{ie}^i$ and base $p_{ib}^i$.}
    \label{fig:pos}
\end{figure}

\begin{figure}
    \centering
    \includegraphics[width=0.33\textwidth]{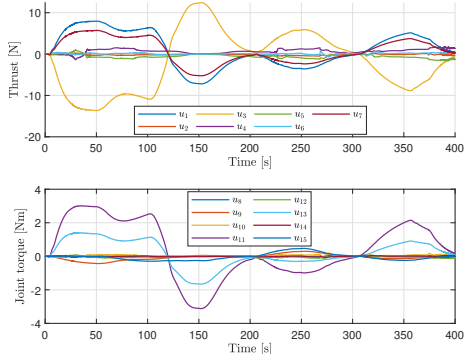}
    \vspace{-1em}
    \caption{The thruster and joint torque control inputs.}
    \label{fig:control-inputs}
\end{figure}

\begin{figure}
    \centering 
    \includegraphics[width=0.33\textwidth]{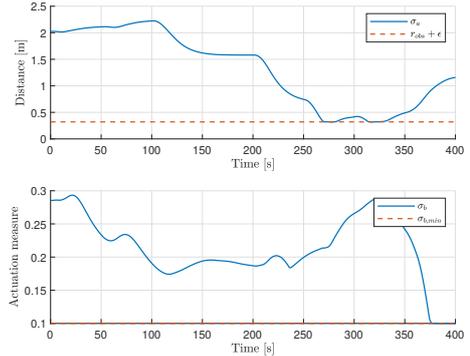}
    \vspace{-1em}
    \caption{The distance to the center of the spherical obstacle $\sigma_a$ and its minimum value $r_{\text{obs}}+\epsilon$, and the actuation measure $\sigma_b$ and its minimum value $\sigma_{b,\text{min}}$.}
    \label{fig:north-west}
\end{figure}

\begin{figure}
    \centering
    \includegraphics[width=0.33\textwidth]{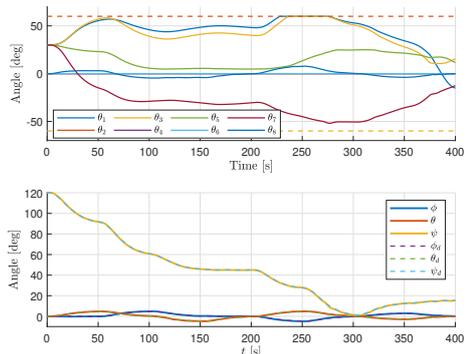}
    \vspace{-1em}
    \caption{The joint angles $\theta$, their maximum and minimum limits $\theta_\text{max}=\SI{60}{\degree}$ and $\theta_\text{min}=\SI{-60}{\degree}$, and the orientation of the end-effector, represented by the roll-pitch-yaw Euler angles $\phi$, $\theta$ and $\psi$.}
    \label{fig:joints-attitude}
\end{figure}

\section{Conclusions and Future Work}\label{sec:conclusion}
This paper has presented a novel task-priority framework for redundancy resolution, dynamic control and control allocation of redundant robotic systems based on a hierarchy of CLF- and CBF-based QPs. The framework provides strict priority, ensuring that lower-priority tasks have no effect on higher-priority tasks, by solving additional QPs to establish distinct priority levels. As a result, lower-priority control objectives can be safely included, without affecting the execution of higher-priority mission-related or safety-related tasks. Additionally, a soft priority measure in the form of slack variables can be utilized in order to prioritize tasks at the same priority level, resulting in considerable design freedom.

% The framework offers considerable design freedom thanks to a soft priority measure in the form of slack variables that help prioritize tasks at the same priority level. Moreover, it provides a strict priority measure, ensuring that lower-priority tasks have no effect on higher-priority tasks, in the form of solving another QP to establish distinct priority levels. As a result, lower-priority control objectives can be safely included, without affecting the execution of higher-priority mission related tasks.

% The proposed framework has been verified in simulation for an AIAUV, which is an overactuated and redundant robotic system. For these types of systems, the framework proposed in this paper solves the redundancy resolution, dynamic control and control allocation problems simultaneously, and thereby ensuring strict priority among tasks at all times. 
The proposed framework has been verified in simulations for an AIAUV, which is an overactuated and redundant robotic system. For these types of systems, the proposed task-priority framework also solves the control allocation problem, which is highly advantageous since control input bounds and rate constraints can be accounted for when resolving redundancy, effectively avoiding a situation in which commanded generalized forces and torques cannot be allocated explicitly, leading to a loss of priority among tasks.
 
Future work is aimed at investigating the robustness of the proposed framework with respect to modeling inaccuracies. This is especially relevant for an underwater vehicle application such as an AIAUV, where accurate identification of the dynamic model parameters is difficult \citep{antonelli2018underwater}. An experimental implementation of the proposed control system on an AIAUV will further validate the framework.

\vspace{-1em}
% \bibliography{ifacconf,bibliography}             % bib file to produce the bibliography
                                                     % with bibtex (preferred)
\bibliography{bibliography}
% \bibliography{bibliography}
                                                   
%\begin{thebibliography}{xx}  % you can also add the bibliography by hand 

%\bibitem[Able(1956)]{Abl:56}
%B.C. Able.
%\newblock Nucleic acid content of microscope.
%\newblock \emph{Nature}, 135:\penalty0 7--9, 1956.

%\bibitem[Able et~al.(1954)Able, Tagg, and Rush]{AbTaRu:54}
%B.C. Able, R.A. Tagg, and M.~Rush.
%\newblock Enzyme-catalyzed cellular transanimations.
%\newblock In A.F. Round, editor, \emph{Advances in Enzymology}, volume~2, pages
%  125--247. Academic Press, New York, 3rd edition, 1954.

%\bibitem[Keohane(1958)]{Keo:58}
%R.~Keohane.
%\newblock \emph{Power and Interdependence: World Politics in Transitions}.
%\newblock Little, Brown \& Co., Boston, 1958.

%\bibitem[Powers(1985)]{Pow:85}
%T.~Powers.
%\newblock Is there a way out?
%\newblock \emph{Harpers}, pages 35--47, June 1985.

%\bibitem[Soukhanov(1992)]{Heritage:92}
%A.~H. Soukhanov, editor.
%\newblock \emph{{The American Heritage. Dictionary of the American Language}}.
%\newblock Houghton Mifflin Company, 1992.

%\end{thebibliography}

% \appendix
% \section{A summary of Latin grammar}    % Each appendix must have a short title.
% \section{Some Latin vocabulary}              % Sections and subsections are supported  
                                                                         % in the appendices.
\end{document}